\documentclass[english,aps,prb,preprint,mathserif,superscriptaddress,preprintnumbers]{revtex4-1}

\usepackage{times}
\usepackage{graphicx}
\usepackage{mathptmx}
\usepackage{lineno}
\usepackage{natmove}
\usepackage{hyperref}
\usepackage{color}
\usepackage{babel}

\begin{document}
\title{Long-lived modulation of plasmonic absorption by ballistic thermal injection}

\author{John A. Tomko}
\affiliation{Department of Materials Science and Engineering, University of Virginia, Charlottesville, VA 22904, USA}

\author{Evan L. Runnerstrom}
\affiliation{Army Research office, CCDC US Army Research Laboratory, Research Triangle Park, NC 27709, USA}
\affiliation{Department of Materials Science and Engineering, North Carolina State University, Raleigh, NC 27695, USA}

\author{Yi-Siang Wang}
\affiliation{Department of Chemistry, University of Southern California, Los Angeles, California 90089, USA}
\affiliation{Department of Physics and Astronomy, University of Southern California, Los Angeles, California 90089, USA}

\author{Joshua R. Nolen}
\affiliation{Institute of Nanoscale Science and Engineering, Vanderbilt University, Nashville, TN 37235, USA}

\author{David H. Olson}
\affiliation{Department of Mechanical and Aerospace Engineering, University of Virginia, Charlottesville, VA 22904, USA}

\author{Kyle P. Kelley}
\affiliation{Department of Materials Science and Engineering, North Carolina State University, Raleigh, NC 27695, USA}

\author{Angela Cleri}
\affiliation{Department of Materials Science and Engineering, Pennsylvania State University, State College, PA USA}

\author{Josh Nordlander}
\affiliation{Department of Materials Science and Engineering, Pennsylvania State University, State College, PA USA}

\author{Joshua D. Caldwell}
\affiliation{Institute of Nanoscale Science and Engineering, Vanderbilt University, Nashville, TN 37235, USA}

\author{Oleg V. Prezhdo}
\affiliation{Department of Chemistry, University of Southern California, Los Angeles, California 90089, USA}
\affiliation{Department of Physics and Astronomy, University of Southern California, Los Angeles, California 90089, USA}

\author{Jon-Paul Maria}
\affiliation{Department of Materials Science and Engineering, North Carolina State University, Raleigh, NC 27695, USA}
\affiliation{Department of Materials Science and Engineering, Pennsylvania State University, State College, PA USA}

\author{Patrick E. Hopkins}
\email{phopkins@virginia.edu}
\affiliation{Department of Materials Science and Engineering, University of Virginia, Charlottesville, VA 22904, USA}
\affiliation{Department of Mechanical and Aerospace Engineering, University of Virginia, Charlottesville, VA 22904, USA}
\affiliation{Department of Physics, University of Virginia, Charlottesville, VA 22904, USA}

\date{\today}
\maketitle


\textbf{Energy and charge transfer across metal-semiconductor interfaces are the fundamental driving forces for a broad range of applications, such as computing, energy harvesting, and photodetection. However, the exact roles and physical separation of these two phenomena remains unclear, particularly in plasmonically-excited systems or cases of strong non-equillibrium. We report on a series of ultrafast plasmonic measurements that provide a direct measure of electronic distributions, both spatially and temporally, following optical excitation of a metal-semiconductor heterostructure. For the first time, we explicitly show that in cases of strong non-equilibrium, a novel energy transduction mechanism arises at the metal/semiconductor interface. We find that hot electrons in the metal contact transfer their energy to pre-existing electrons in the semiconductor, without transfer of charge. These experimental results findings are well-supported by both rigorous multilayer optical modeling and first-principle, \textit{ab initio} calculations.}


\section*{Introduction}

Light-matter interactions that induce charge and energy transfer across interfaces form the foundation for multiple technological phenomena including photocatalysis \cite{Serpone2012,Kim2015}, energy harvesting \cite{Clavero2014}, and photodetection \cite{Knight2011}. One of the most common mechanisms associated with these processes relies on injection of the carrier itself; this electron injection is one of the most common mechanisms associated with these processes and is considered to be a relatively well-understood phenomenon that can be manipulated at nearly all length (nm to bulk) and time (sub-ps to steady-state) scales  \cite{Ji2017,Hong2014}. However, the exact role of electron injection in these applications remains unclear. Plasmon-assisted photocatalytic efficiencies can improve when intermediate insulation layers are used to inhibit charge transfer \cite{Asapu2019,Chen2012} or when off-resonance excitation is used \cite{Priebe2013}, yet the typical assumption is that charge injection is responsible for the catalytic enhancement \cite{Narang2016,Brongersma2015}. While considerable effort has been paid to characterizing novel plasmon-induced behavior and transport mechanisms \cite{Sundararaman2014, Wu2015}, current understanding still relies on a charge injection motif that has been established through long-time, quasi-steady state experiments. This is despite the fact that photothermal effects play an explicit role in such experiments when charge transfer is inhibited, and can further behave in a synergistic manner even when electron injection is `allowed' to occur \cite{Yu2018}.

We hypothesize a novel energy transduction mechanism that arises from the non-equilibrium dynamics of excited charges at metal-semiconductor interfaces, without charge injection. Such an energy transduction mechanism has yet to be demonstrated. These non-equilibrium dynamics exist when electron temperature is highly elevated relative to the crystal lattice temperature and can only occur at ultra-fast timescales. In this regime, interfacial energy injection (without concomitant charge injection) from an excited metal to a non-metal offers the possibliity of remotely manipulating the photonic and electronic properties of non-metals without relying on specific photonic or electronic excitations in the non-metal or metal contact. This would enable remote control of the non-metal's functional properties via injected heat, a mechanism of modulation that relies only on the 2nd Law of Thermodynamics.

To test this hypothesis, we must identify a system that lacks facile charge injection across a metal-semiconductor interface when in an excited state. Electron injection can only occur in response to a gradient in the electronic chemical potential or Fermi level induced by an extrinsic perturbation (e.g., photonic excitation, temperature differences, or applied bias). Additionally, the excited carriers must have enough energy to traverse or tunnel through any Schottky barrier that exists at the interface. But what happens when the excited carriers are not accompanied by a change in electronic chemical potential? What if a Schottky barrier does not exist at the interface? We address the first question by using a noble metal, gold, with a nearly temperature-independent electronic chemical potential. We address the second question by using a degenerate semiconductor, yttrium-doped cadmium oxide (Y:CdO), which forms a barrier-free ohmic contact with gold. In concert, these two effects should eliminate the ability for light-induced charge transfer to occur at this metal-semiconductor interface. This opens the path to an entirely new means of interfacial energy transduction: excited electrons in the metal ballistically couple their energy, rather than charge, to the pre-existing electrons of the semiconductor. This ballistic thermal injection (BTI) process has yet to be recognized as a means of energy transfer as it is very elusive to observation: the process is contingent on non-equilibrium conditions to facilitate energy injection across an interface, yet simultaneously requires the absence of a potential gradient to limit charge injection. In the case of continuous wave, constant irradiation experiments, where the electrons and lattice are approximately in thermal equilibrium, this BTI process would be indistinguishable from charge injection. This can be alleviated by using ultra-fast, spectrally resolved optical pump-probe experiments to look for spectroscopic signatures that would indicate alternative relaxation pathways \cite{Kozawa2016}. In this work, we investigate the non-equilibrium energy transfer across a metal/semiconductor interface with a novel ultrafast metrology that temporally and spectrally resolves changes in the optical properties of the semiconductor at and around the epsilon-near-zero (ENZ) condition. Because the optical properties of an ENZ medium are highly sensitive to spatial variations in electron density and energy  \cite{Feigenbaum2010,Folland2019,Tandon2019}, this technique provides unique insight into the energy transduction and relaxation mechanisms within the semiconductor. Furthermore, as we optically excite the metal contact from the air/metal interface with a pump pulse, the possibility for plasmonic resonance conditions to induce energy transfer is eliminated. Rather, we monitor the ENZ condition of the doped cadmium oxide semiconductor with a sub-picosecond probe pulse following excitation; this ENZ mode is a bulk plasmon, and functions as a sensor of charge and energy transfer from the metal into the CdO.

\section*{Materials system}

Gold in contact with Y:CdO is an ideal metal-semiconductor heterostructure to understand and manipulate our proposed non-equilibrium driven energy transfer pathway. Due to its large electron affinity (5.9 eV), CdO (Y:CdO) is intrinsically (extrinsically) an n-type degenerate semiconductor with a Fermi level residing deep within the conduction band. Y:CdO is thus expected to always form an ohmic contact with gold (work function $\sim$5.4 eV). Simultaneously, below an electron temperature of $\sim$3000 K, the Fermi level or electronic chemical potential of gold remains relatively constant, which limits thermally driven electronic diffusion \cite{Lin2008}.

With these two factors in mind, charge carriers are not expected to flow between two materials following pulsed excitation of conduction electrons in gold. Importantly, the weak electron-phonon coupling factor of Au allows its excited electrons to stay at highly elevated temperatures for prolonged times relative to other metals. Thus, following ultrafast excitation of Au, energy can ballistically traverse the Au film through the electron subsystem to reach the Au/CdO interface, and this can occur before the electron subsystem thermalizes and loses its excess energy to the Au phonon subsystem/lattice. As this energy front in Au reaches the Au/CdO interface, the hot electron subsystem in Au can then directly couple excess energy into CdO's electron subsystem due to the strong overlap of electronic wavefunctions between the two materials. This results in ballistic energy transduction across the interface without charge transfer. A schematic of the typically-assumed charge injection process and our proposed BTI process are shown in Fig.~1a and 1b, respectively. We note that the scattering processes following either charge injection or the proposed BTI energy transfer mechanism, such as hot electron-electron scattering, are excluded for clarity.

In addition to the unique ability to separate energy and charge transfer processes, doped CdO is a model mid-infrared plasmonic material that supports free electron densities on the order of $10^{19}$-$10^{21}$ cm$^{-3}$ while maintaining high electron mobilities of 300-500 cm$^{-2}$ V$^{-1}$ s$^{-1}$ \cite{Sachet2015,Runnerstrom2017,Runnerstrom2019}. These electronic properties enable strong, sharp, and resonant light-matter interactions at mid- to near-infrared frequencies. As in other plasmonic materials, these resonances are highly sensitive to local changes in the electronic environment, including electron density, effective mass, and dielectric constant. A key difference here is that, rather than probing surface and environment-sensitive surface plasmon polaritons as is common in thin-film metallic plasmonics, we optically monitor the resonant radiative bulk ENZ mode (also known as the Brewster mode) of the CdO following optical excitation of the Au film. Our sample geometry is especially convenient for this experiment, as the Au film also acts as a mirror that enhances the free-space coupling of the ENZ/Brewster mode, which allows us to monitor the reflectivity of the CdO/Au heterostructure from the backside through a transparent substrate. Thanks to the high electronic mobility, the ENZ/Brewster mode manifests as a sharp resonant dip in the reflectivity (i.e., an absorption peak), making it straightforward to resolve changes in optical behavior. This scheme provides high sensitivity to the spatial distribution of electronic energy in the semiconductor as well as its temporal evolution following energy transfer from the metallic film; a schematic of this measurement technique and the sample configuration is shown in Fig.~1c while greater detail can be found in the Supporting Information. In the following, we demonstrate long-lived modulation of the CdO ENZ mode following Au excitation, and show that this modulation can be explained only by the BTI process and not by a charge transfer mechanism.

\section*{Excited-carrier dynamics at the Au/CdO interface}
To gain initial insight into the hot-electron dynamics at the Au/CdO interface, and ensure our hypothesis of ballistic thermal injection can occur in these systems, we first perform time-domain thermoreflectance (TDTR) measurements \cite{Choi2014,Radue2018} on the Au/CdO samples supported by a sapphire (Al\textsubscript{2}O\textsubscript{3}) substrate, where both the pump and probe are focused on top Au film/air interface. These measurements allow us to quantify ultra-fast energy flow across the Au/CdO and CdO/substrate interfaces and asseess time scales of thermal transport at each interface. Our TDTR experiment can be conceptually summarized as follows. First, the sub-ps pulse excites electronic carriers in the metal, which quickly (10's of fs) scatter and decay into a `hot' Fermi-Dirac distribution. At this stage, the electronic subsystem is at a greatly elevated temperature, while the phononic subsystem remains cold (i.e., approximately at its initial temperature). In the case of metals in contact with dielectrics (e.g., Al$_2$O$_3$), these hot carriers are spatially confined to the metal and can neither traverse the metal/dielectric interface nor directly couple their energy to electrons or phonons in the insulator. Instead, they couple to and lose energy to the phononic subsystem (i.e., the lattice) of the metal film. Following this electron-phonon coupling, phonon-based thermal conduction out of the metal and into the insulator over hundreds of picoseconds to nanoseconds. The entire process is monitored via temporal changes in the metal's thermoreflectance: the reflectance signal decays with time as the surface of the metal film loses energy to the underlying substrate via phonon interactions at the metal/non-metal interface \cite{Qiu1994a,Qiu1994,Lyeo2006,Giri2019}.

Conversely, when Au is are in direct contact with a different metal (e.g., Au/Ti or Au/Pt interfaces \cite{Yu2019,Ratchford2017,Giri2015,Choi2014}, the excited electrons are no longer confined and can directly couple energy across the metal/metal interface, leading to an `instantaneous' (sub-picosecond) temperature rise of electrons in the underlying metal. These hot-electrons in the underlying metal can subsequently lose energy to the phononic subsystem more quickly than those in the Au, which creates a sub-surface temperature rise. This sub-surface heating leads to a bi-directional heat flux originating from some depth under the Au/air surface, and manifests as a temporally delayed rise in the thermoreflectance signal as heat is conducted back into the Au through the lattice (e.g., `back-heating'). This back-heating signature is thus a strong indication of electron energy transfer from the Au film into the underlying metal film \cite{Choi2014}. This phenomenon acts as a transient thermal diode: the injection of energy from Au to the buried metal substrate occurs on sub-picosecond time-scales, whereas the lattice-mediated `back-heating' occurs on hundreds of picoseconds to nanoseconds.

Indeed, TDTR measurements (Fig.~2a) on 15 nm Au/100 nm CdO films display the back-heating and the transient thermal diode effects, indicating that hot electrons in the Au are transferring their energy to free electrons in the CdO at ultra-fast time scales. Conversely, when a thin 15 nm dielectric HfO\textsubscript{2} layer is placed between the Au and CdO, this energy transfer process is inhibited and no sign of back-heating is observed. This observation rules out any possibility that the pump beam is directly exciting the CdO film and is the cause for subsurface heating, as the HfO$_{2}$ is optically transparent to the pump wavelength. It also confirms our posit that a dielectric barrier, which limits the electron wavefunction overlap between Au and CdO, will inhibit thermal energy transduction across the Au/CdO interface.

As mentioned, interpreting of spectroscopic characteristics in typical pump-probe experiments such as TDTR can be quite difficult, as several mechanisms can lead to nearly identical signatures. In both metal/metal and Au/CdO heterostructures, there are two potential mechanisms for subsurface heating at ultra-fast time scales: charge injection and our proposed BTI process. In both situations, ohmic contact between two materials with high carrier density may lead one to expect facile charge injection into the underlying metal or CdO. In that case, the injected electron would eventually decay within the underlying metal or CdO via electron-phonon coupling and induce a sub-surface temperature rise. Contrarily, the back-heating signatures could result from BTI without any concomitant charge flow: optically deposited energy ballistically traverses Au's electronic subsystem to reach the interface, where it couples directly to the electronic subsystem of the underlying metal or CdO. Following this BTI process, the now-excited electrons in the underlying metal/CdO would couple to the lattice and lead to sub-surface heating.

To distinguish these processes, we simulate the electronic interactions involved in the nonequilbrium photo-induced dynamics via \textit{ab initio} real-time time-dependent density functional theory (TDDFT) for electrons coupled to nonadiabatic molecular dynamics (NAMD) for atomic motions (Fig.~2c). The calculations reveal that following photo-excitation of the Au, the hot electron remains within Au, with the tail of its wavefunction extending into the CdO layer. This tail directly couples energy to electrons in the CdO within picoseconds, followed by electron-phonon relaxation through high-frequency phonon modes \cite{Zhou2019}. Concurrently, (Fig.~2b and 2c) the hot electron's wave function quickly ($\sim$6 ps) re-localizes within the Au, which quenches BTI. The calculations show that energy transfer from Au to CdO is mediated by hot electron energy coupling and not a phononic or charge-injection process. This result, obtained with a small representation of the Au/CdO interface, agrees well with our observed experimental trends and supports the two-temperature interpretation of our results.

\section*{Remote modulation of the ENZ mode}
This novel BTI mechanism offers a unique opportunity to manipulate the electrons in an optically active material without relying on direct photonic or electronic perturbations, as is commonly done. Instead, the BTI process remotely manipulates the photonic or electronic response of a non-metal through optical excitation of the metal transducer. To demonstrate this, we perform additional pump-probe experiments on the heterostructure, where a tunable wavelength probe monitors the ENZ behavior of the CdO, following visible (520 nm) excitation of the Au film, at picosecond time scales. This ENZ mode is highly sensitive to the electronic environment and free carrier dynamics within the CdO. By monitoring CdO's optical behavior, we gain direct knowledge of how the electronic environment changes with time during the BTI process. See supporting information for additional details on ENZ modes and this experiment.

Following the pump excitation of gold, the ENZ absorption peak of CdO at $\sim$3800 nm red-shifts, which increases absorption (decreases reflectivity) at longer wavelengths. Concurrently, absorption (reflectivity) decreases (increases) at shorter wavelengths, though to a lesser extent. This asymmetric red-shift persists for hundreds of picoseconds following thermal excitation of the gold film.

Our observation of a red-shift immediately negates the possibility of electron injection from the Au into the CdO layer. Based on the Drude model, the ENZ mode of a conducting thin film is centered at the screened plasma frequency, $\omega_{ENZ} = \omega_{p}/\sqrt{\epsilon_{\infty}}$, where $\epsilon_{\infty}$ is the high-frequency dielectric constant, $\omega_{p} = \sqrt{n_{e}e^{2}/m^{*}\epsilon_{0}}$, and where $n_{e}$ is the free electron density and $m^{*}$ is the effective mass of the electrons. Therefore, any electron injection would increase $n_{e}$ and result in a blue-shift of the ENZ absorption peak, in direct contrast to our experimental results.

While one could argue that hole injection or heating of the CdO layer may be occurring, which would result in a decrease in $n_{e}$ and an increase in $m^{*}$, respectively, both would lead to a symmetric red-shift. As indicated prior and shown in Fig.~3a and Fig.~3c, we clearly observe an \textit{asymmetric} red-shift. To understand the origin of this asymmetry, we perform optical transfer matrix method (TMM, Fig.~3b,c) calculations to model how the electron distribution changes in the CdO layer following Au excitation. (A more detailed description can be found in the supporting information.)

Our TMM calculations reveal that $n_{e}$ remains unchanged in the CdO, regardless of pump fluence or relative pump-probe time-delay indicating that neither electron nor hole injection occurs. Rather, in accordance with our BTI hypothesis, the energy distribution of these charges changes based on the following physical processes. First, the Au couples its electronic energy to the CdO, thus heating free electrons within a thin ($<$5 nm) CdO layer near the Au/CdO interface. This energy heats the electrons, which leads to a local increase in $m^{*}$ within this layer \cite{Yang2017}. Additionally, because scattering rate is inversely proportional to the effective mass, our model also captures a local decrease in the damping frequency of the free electrons. As this heated layer is now out-of-equilibrium relative to the remainder of the CdO, electrons from the bulk of the film diffuse towards the hot CdO layer, which temporarily forms an accumulation layer near the Au/CdO interface and slightly decreases $n_{e}$ throughout the remainder of the CdO film. This increase in the local number density leads to a corresponding decrease in the high frequency permittivity, $\epsilon_{\infty}$ \cite{Liu2016,Nolen2019}. As shown in Fig.~3c, the TMM calculations only re-produce our experimental results, with excellent agreement, when our proposed BTI energy transfer mechanism is invoked. TMM-simulated electron or hole injection cannot explain the observed asymmetric red-shift in ENZ absorption, regardless of whether the charge is injected into the full CdO layer or a thin slab near the interface. These TMM simulations incorporate rigorous relationships between the optical parameters (plasma frequency, scattering rate, high-frequency dielectric constant) and electronic parameters (electron number density, mobility, and effective mass), consistent with previous observations \cite{Liu2016,Nolen2019}.

Finally, we note that electronic non-equilibrium within the CdO, and thus associated remote modulation of the ENZ mode, is altered for prolonged times with a 1/e decay time of approximately 700 ps (Fig.~3d). This length of modulation is orders of magnitude greater than previous works investigating light-induced changes in plasmonic absorption of the CdO film. This long-lived modulation is ultimately enabled by the aforementioned `transient thermal diode' effect that is enabled by the BTI process across the Au/CdO interface.

\section*{Conclusion}
We find that under non-equilibrium conditions, photo-excited metals can undergo an electron-mediated ballistic energy transfer process that can remotely modulate the electronic environment of an underlying degenerate semiconductor. We term this effect `ballistic thermal injection' (BTI) to distinguish it from the well-studied charge injection effects caused by hot electrons traversing an energy barrier. Our results are explained through transient reflectivity (TDTR) measurements to understand the fundamental carrier dynamics at the Au/CdO interface. Additionally, we perform mid-infrared, ultra-fast pump-probe experiments to obtain spectral, temporal, and ultimately spatial resolution of the electronic environment within the CdO layer following excitation of the Au film. Ultimately, the carrier dynamics derived from the transient spectra fully agree with our TDTR results, and can only be accurately modeled by the BTI process rather than by charge injection. From an application perspective, although BTI does not lend itself to ultra-fast switching, one could easily conceive of several applications where longer modulation lifetimes are more beneficial, such as catalysis, sensing, or energy harvesting, through an increased `interaction’ time. The long-lived modulation of epsilon-near-zero media such, in particular, can be utilized as a selective emitter for thermophotovoltaic (TPV) conversion, as BTI can prolong the desired state of emissivity, or nonlinear phenomena such as four-wave mixing and high harmonic generation. Furthermore, the proposed BTI energy transduction mechanism could provide explanation for discrepancies between current theory and experimental results in light-matter interactions, such as photocatalysis, that rely on either the framework of a charge transfer motif or solely phononic heat conduction.

\section*{Methods}
\subsection{Sample fabrication}
Heteroepitaxial thin films of Y-doped CdO were prepared by reactive HiPIMS from a metallic Cd target and by reactive RF sputtering from a metallic Y target. All the films were grown on epitaxial-polished Al$_{2}$O$_{3}$. Film thicknesses were determined by X-ray reflectivity measurements. Finally, 15 nm of Au was deposited via electron-beam evaporation.

\subsection{Time domain thermoreflectance}
To gain insight to the temperature response of the Au film following pulsed excitation, we performed and repeated TDTR measurements on two different systems; these are identical in all aspects, with the difference being that one is two-tint (e.g., the pump and probe are both centered at a wavelength of 800 nm), while the other system is two-color (e.g., the pump is frequency doubled via a BBO crystal to a wavelength of 400 nm and the probe operates at the fundamental 800 nm output). In both cases, the pump beam is modulated at 8.8 MHz and focused to the sample surface

\subsection{Optical pump-infrared probe measurements}
To monitor the epsilon-near-zero (ENZ) condition of the doped CdO film following excitation of the Au film, we utilize a second pump-probe system to measure the differential reflectivity. This experiment relies on a 400 femtosecond laser operating at its fundamental wavelength of 1040 nm with a repetition rate of 500 kHz (Spectra Physics 30W Spirit-HP). The output is split into two paths; the first path is frequency-doubled to a wavelength of 520 nm, which operates as the pump pulse. This pump passes down a mechanical delay stage, mechanically chopped at a frequency of 451 Hz, and then focused to the surface of the Au metal film at the Au/air interface. The other portion of the beam is passed through an optical parametric amplifier, where the signal beam is filtered out and the idler operates as the probe; this probe beam is spectrally varied in 10 nm increments from 3400-4200 nm and focused onto the Au surface at the CdO/Au interface. The sapphire/100 nm CdO/Au sample is mounted on a mechanical goniometer to allow for high-precision in defining the incident angle of the probe beam. Although the ENZ resonance for this structure has minimal angular-dependence and inhibits reflection at the plasma frequency from $~30$ to $~70$ degrees, the peak absorption occurs at $~55$ degrees and is thus where we perform the majority of our experiments. Note, the probe beam is p-polarized for the majority of our experimental results as only this polarization is sensitive to the ENZ mode.

\subsection{Time-dependent density functional theory (TDDFT)}
The nonequilibrium dynamics of coupled electrons and nuclei are simulated by real-time TDDFT for the electronic evolution coupled
to ionic motions via NAMD.
The simulation cell is formed from five layers of CdO and three layers of Au atoms, to imitate the experimental thickness ratio of the CdO and Au films, and 20\AA of vaccum separating the slabs. The simulation cell contains 138 atoms total, including 60 Cd, 60 O and 18 Au atoms. The geometric and electronic structure calculations, and adiabatic molecular dynamics (MD) are carried out using the Quantum Espresso \cite{Giannozzi2009,Gianozzi2017}. The simulations use the PBE functional with Hubbard parameters 12 eV for Cd and 8 eV for O, taken from \cite{Aras2014,Goh2017} in order to obtain the proper alignment of the Au Fermi level and the CdO band edges. The plane-wave basis set energy cutoﬀ is set to 820 eV. A 4$\times$6$\times$1 k-point mesh grid is used.

After the structure optimization, the system is brought up to 300 K in the canonical ensemble using the Andersen thermostat. Then, a 5 ps trajectory is obtained in the microcanonical ensemble and used for the subsequent NAMD simulations. The adiabatic state energies and NA are calculated for each step of the MD run. The 5 ps NA Hamiltonian obtained this way is iterated several times in order to simulate longer time dynamics. The nuclear dynamics of the system is rather simple, involving no long term atomic rearrangements, and therefore 5 ps are sufficient to sample nuclear motions that drive the electron dynamics. The NAMD calculations are carried out using the quantum-classical fewest switches surface hopping (FSSH) technique implemented with real-time TDDFT in the PYXAID code \cite{Akimov2013,Akimov2014}.  1000 initial geometries are sampled from the 5 ps MD trajectory, and 100 stochastic realizations of the FSSH process are generated for each geometry. Additional details on our TDDFT calculations can be found in the Supporting Information.

\bibliography{AuCdOBib}

\begin{thebibliography}{41}%
\makeatletter
\providecommand \@ifxundefined [1]{%
 \@ifx{#1\undefined}
}%
\providecommand \@ifnum [1]{%
 \ifnum #1\expandafter \@firstoftwo
 \else \expandafter \@secondoftwo
 \fi
}%
\providecommand \@ifx [1]{%
 \ifx #1\expandafter \@firstoftwo
 \else \expandafter \@secondoftwo
 \fi
}%
\providecommand \natexlab [1]{#1}%
\providecommand \enquote  [1]{``#1''}%
\providecommand \bibnamefont  [1]{#1}%
\providecommand \bibfnamefont [1]{#1}%
\providecommand \citenamefont [1]{#1}%
\providecommand \href@noop [0]{\@secondoftwo}%
\providecommand \href [0]{\begingroup \@sanitize@url \@href}%
\providecommand \@href[1]{\@@startlink{#1}\@@href}%
\providecommand \@@href[1]{\endgroup#1\@@endlink}%
\providecommand \@sanitize@url [0]{\catcode `\\12\catcode `\$12\catcode
  `\&12\catcode `\#12\catcode `\^12\catcode `\_12\catcode `\%12\relax}%
\providecommand \@@startlink[1]{}%
\providecommand \@@endlink[0]{}%
\providecommand \url  [0]{\begingroup\@sanitize@url \@url }%
\providecommand \@url [1]{\endgroup\@href {#1}{\urlprefix }}%
\providecommand \urlprefix  [0]{URL }%
\providecommand \Eprint [0]{\href }%
\providecommand \doibase [0]{http://dx.doi.org/}%
\providecommand \selectlanguage [0]{\@gobble}%
\providecommand \bibinfo  [0]{\@secondoftwo}%
\providecommand \bibfield  [0]{\@secondoftwo}%
\providecommand \translation [1]{[#1]}%
\providecommand \BibitemOpen [0]{}%
\providecommand \bibitemStop [0]{}%
\providecommand \bibitemNoStop [0]{.\EOS\space}%
\providecommand \EOS [0]{\spacefactor3000\relax}%
\providecommand \BibitemShut  [1]{\csname bibitem#1\endcsname}%
\let\auto@bib@innerbib\@empty
\bibitem [{\citenamefont {Serpone}\ and\ \citenamefont
  {Emeline}(2012)}]{Serpone2012}%
  \BibitemOpen
  \bibfield  {author} {\bibinfo {author} {\bibfnamefont {N.}~\bibnamefont
  {Serpone}}\ and\ \bibinfo {author} {\bibfnamefont {A.~V.}\ \bibnamefont
  {Emeline}},\ }\href {\doibase 10.1021/jz300071j} {\bibfield  {journal}
  {\bibinfo  {journal} {The Journal of Physical Chemistry Letters}\ }\textbf
  {\bibinfo {volume} {3}},\ \bibinfo {pages} {673} (\bibinfo {year}
  {2012})}\BibitemShut {NoStop}%
\bibitem [{\citenamefont {Kim}\ \emph {et~al.}(2015)\citenamefont {Kim},
  \citenamefont {Hyosun},\ and\ \citenamefont {Park}}]{Kim2015}%
  \BibitemOpen
  \bibfield  {author} {\bibinfo {author} {\bibfnamefont {S.~M.}\ \bibnamefont
  {Kim}}, \bibinfo {author} {\bibfnamefont {L.}~\bibnamefont {Hyosun}}, \ and\
  \bibinfo {author} {\bibfnamefont {J.~Y.}\ \bibnamefont {Park}},\ }\href
  {\doibase 10.1007/s10562-014-1418-y} {\bibfield  {journal} {\bibinfo
  {journal} {Catalysis Letters}\ }\textbf {\bibinfo {volume} {145}},\ \bibinfo
  {pages} {299} (\bibinfo {year} {2015})}\BibitemShut {NoStop}%
\bibitem [{\citenamefont {Clavero}(2014)}]{Clavero2014}%
  \BibitemOpen
  \bibfield  {author} {\bibinfo {author} {\bibfnamefont {C.}~\bibnamefont
  {Clavero}},\ }\href {\doibase 10.1038/nphoton.2013.238} {\bibfield  {journal}
  {\bibinfo  {journal} {Nature Photonics}\ }\textbf {\bibinfo {volume} {8}},\
  \bibinfo {pages} {95} (\bibinfo {year} {2014})}\BibitemShut {NoStop}%
\bibitem [{\citenamefont {Knight}\ \emph {et~al.}(2011)\citenamefont {Knight},
  \citenamefont {Sobhani}, \citenamefont {Nordlander},\ and\ \citenamefont
  {Halas}}]{Knight2011}%
  \BibitemOpen
  \bibfield  {author} {\bibinfo {author} {\bibfnamefont {M.~W.}\ \bibnamefont
  {Knight}}, \bibinfo {author} {\bibfnamefont {H.}~\bibnamefont {Sobhani}},
  \bibinfo {author} {\bibfnamefont {P.}~\bibnamefont {Nordlander}}, \ and\
  \bibinfo {author} {\bibfnamefont {N.~J.}\ \bibnamefont {Halas}},\ }\href@noop
  {} {\bibfield  {journal} {\bibinfo  {journal} {Science}\ }\textbf {\bibinfo
  {volume} {332}},\ \bibinfo {pages} {702} (\bibinfo {year}
  {2011})}\BibitemShut {NoStop}%
\bibitem [{\citenamefont {Ji}\ \emph {et~al.}(2017)\citenamefont {Ji},
  \citenamefont {Hong}, \citenamefont {Zhang}, \citenamefont {Zhang},
  \citenamefont {Huang}, \citenamefont {Cao}, \citenamefont {Qiao},
  \citenamefont {Liu}, \citenamefont {Liang}, \citenamefont {Jin},
  \citenamefont {Jiao}, \citenamefont {Shi}, \citenamefont {Meng},\ and\
  \citenamefont {Liu}}]{Ji2017}%
  \BibitemOpen
  \bibfield  {author} {\bibinfo {author} {\bibfnamefont {Z.}~\bibnamefont
  {Ji}}, \bibinfo {author} {\bibfnamefont {H.}~\bibnamefont {Hong}}, \bibinfo
  {author} {\bibfnamefont {J.}~\bibnamefont {Zhang}}, \bibinfo {author}
  {\bibfnamefont {Q.}~\bibnamefont {Zhang}}, \bibinfo {author} {\bibfnamefont
  {W.}~\bibnamefont {Huang}}, \bibinfo {author} {\bibfnamefont
  {T.}~\bibnamefont {Cao}}, \bibinfo {author} {\bibfnamefont {R.}~\bibnamefont
  {Qiao}}, \bibinfo {author} {\bibfnamefont {C.}~\bibnamefont {Liu}}, \bibinfo
  {author} {\bibfnamefont {J.}~\bibnamefont {Liang}}, \bibinfo {author}
  {\bibfnamefont {C.}~\bibnamefont {Jin}}, \bibinfo {author} {\bibfnamefont
  {L.}~\bibnamefont {Jiao}}, \bibinfo {author} {\bibfnamefont {K.}~\bibnamefont
  {Shi}}, \bibinfo {author} {\bibfnamefont {S.}~\bibnamefont {Meng}}, \ and\
  \bibinfo {author} {\bibfnamefont {K.}~\bibnamefont {Liu}},\ }\href {\doibase
  10.1021/acsnano.7b04541} {\bibfield  {journal} {\bibinfo  {journal} {ACS
  Nano}\ }\textbf {\bibinfo {volume} {11}},\ \bibinfo {pages} {12020} (\bibinfo
  {year} {2017})}\BibitemShut {NoStop}%
\bibitem [{\citenamefont {Hong}\ \emph {et~al.}(2014)\citenamefont {Hong},
  \citenamefont {Kim}, \citenamefont {Shi}, \citenamefont {Zhang},
  \citenamefont {Jin}, \citenamefont {Sun}, \citenamefont {Tongay},
  \citenamefont {Wu}, \citenamefont {Zhang},\ and\ \citenamefont
  {Wang}}]{Hong2014}%
  \BibitemOpen
  \bibfield  {author} {\bibinfo {author} {\bibfnamefont {X.}~\bibnamefont
  {Hong}}, \bibinfo {author} {\bibfnamefont {J.}~\bibnamefont {Kim}}, \bibinfo
  {author} {\bibfnamefont {S.~F.}\ \bibnamefont {Shi}}, \bibinfo {author}
  {\bibfnamefont {Y.}~\bibnamefont {Zhang}}, \bibinfo {author} {\bibfnamefont
  {C.}~\bibnamefont {Jin}}, \bibinfo {author} {\bibfnamefont {Y.}~\bibnamefont
  {Sun}}, \bibinfo {author} {\bibfnamefont {S.}~\bibnamefont {Tongay}},
  \bibinfo {author} {\bibfnamefont {J.}~\bibnamefont {Wu}}, \bibinfo {author}
  {\bibfnamefont {Y.}~\bibnamefont {Zhang}}, \ and\ \bibinfo {author}
  {\bibfnamefont {F.}~\bibnamefont {Wang}},\ }\href {\doibase
  10.1038/nnano.2014.167} {\bibfield  {journal} {\bibinfo  {journal} {Nature
  Nanotechnology}\ }\textbf {\bibinfo {volume} {9}},\ \bibinfo {pages} {682}
  (\bibinfo {year} {2014})}\BibitemShut {NoStop}%
\bibitem [{\citenamefont {Asapu}\ \emph {et~al.}(2019)\citenamefont {Asapu},
  \citenamefont {Claes}, \citenamefont {Ciocarlan}, \citenamefont {Minjauw},
  \citenamefont {Detavernier}, \citenamefont {Cool}, \citenamefont {Bals},\
  and\ \citenamefont {Verbruggen}}]{Asapu2019}%
  \BibitemOpen
  \bibfield  {author} {\bibinfo {author} {\bibfnamefont {R.}~\bibnamefont
  {Asapu}}, \bibinfo {author} {\bibfnamefont {N.}~\bibnamefont {Claes}},
  \bibinfo {author} {\bibfnamefont {R.-G.}\ \bibnamefont {Ciocarlan}}, \bibinfo
  {author} {\bibfnamefont {M.}~\bibnamefont {Minjauw}}, \bibinfo {author}
  {\bibfnamefont {C.}~\bibnamefont {Detavernier}}, \bibinfo {author}
  {\bibfnamefont {P.}~\bibnamefont {Cool}}, \bibinfo {author} {\bibfnamefont
  {S.}~\bibnamefont {Bals}}, \ and\ \bibinfo {author} {\bibfnamefont {S.~W.}\
  \bibnamefont {Verbruggen}},\ }\href {\doibase 10.1021/acsanm.9b00485}
  {\bibfield  {journal} {\bibinfo  {journal} {ACS Applied Nano Materials}\
  }\textbf {\bibinfo {volume} {2}},\ \bibinfo {pages} {4067} (\bibinfo {year}
  {2019})}\BibitemShut {NoStop}%
\bibitem [{\citenamefont {Chen}\ \emph {et~al.}(2012)\citenamefont {Chen},
  \citenamefont {Wu}, \citenamefont {Wu},\ and\ \citenamefont
  {Tsai}}]{Chen2012}%
  \BibitemOpen
  \bibfield  {author} {\bibinfo {author} {\bibfnamefont {J.~J.}\ \bibnamefont
  {Chen}}, \bibinfo {author} {\bibfnamefont {J.~C.}\ \bibnamefont {Wu}},
  \bibinfo {author} {\bibfnamefont {P.~C.}\ \bibnamefont {Wu}}, \ and\ \bibinfo
  {author} {\bibfnamefont {D.~P.}\ \bibnamefont {Tsai}},\ }\href {\doibase
  10.1021/jp309901y} {\bibfield  {journal} {\bibinfo  {journal} {Journal of
  Physical Chemistry C}\ }\textbf {\bibinfo {volume} {116}},\ \bibinfo {pages}
  {26535} (\bibinfo {year} {2012})}\BibitemShut {NoStop}%
\bibitem [{\citenamefont {Priebe}\ \emph {et~al.}(2013)\citenamefont {Priebe},
  \citenamefont {Karnahl}, \citenamefont {Junge}, \citenamefont {Beller},
  \citenamefont {Hollmann},\ and\ \citenamefont {Br{\"{u}}ckner}}]{Priebe2013}%
  \BibitemOpen
  \bibfield  {author} {\bibinfo {author} {\bibfnamefont {J.~B.}\ \bibnamefont
  {Priebe}}, \bibinfo {author} {\bibfnamefont {M.}~\bibnamefont {Karnahl}},
  \bibinfo {author} {\bibfnamefont {H.}~\bibnamefont {Junge}}, \bibinfo
  {author} {\bibfnamefont {M.}~\bibnamefont {Beller}}, \bibinfo {author}
  {\bibfnamefont {D.}~\bibnamefont {Hollmann}}, \ and\ \bibinfo {author}
  {\bibfnamefont {A.}~\bibnamefont {Br{\"{u}}ckner}},\ }\href {\doibase
  10.1002/anie.201306504} {\bibfield  {journal} {\bibinfo  {journal}
  {Angewandte Chemie - International Edition}\ }\textbf {\bibinfo {volume}
  {52}},\ \bibinfo {pages} {11420} (\bibinfo {year} {2013})}\BibitemShut
  {NoStop}%
\bibitem [{\citenamefont {Narang}\ \emph {et~al.}(2016)\citenamefont {Narang},
  \citenamefont {Sundararaman},\ and\ \citenamefont {Atwater}}]{Narang2016}%
  \BibitemOpen
  \bibfield  {author} {\bibinfo {author} {\bibfnamefont {P.}~\bibnamefont
  {Narang}}, \bibinfo {author} {\bibfnamefont {R.}~\bibnamefont
  {Sundararaman}}, \ and\ \bibinfo {author} {\bibfnamefont {H.~A.}\
  \bibnamefont {Atwater}},\ }\href {\doibase 10.1515/nanoph-2016-0007}
  {\enquote {\bibinfo {title} {{Plasmonic hot carrier dynamics in solid-state
  and chemical systems for energy conversion}},}\ } (\bibinfo {year}
  {2016})\BibitemShut {NoStop}%
\bibitem [{\citenamefont {Brongersma}\ \emph {et~al.}(2015)\citenamefont
  {Brongersma}, \citenamefont {Halas},\ and\ \citenamefont
  {Nordlander}}]{Brongersma2015}%
  \BibitemOpen
  \bibfield  {author} {\bibinfo {author} {\bibfnamefont {M.~L.}\ \bibnamefont
  {Brongersma}}, \bibinfo {author} {\bibfnamefont {N.~J.}\ \bibnamefont
  {Halas}}, \ and\ \bibinfo {author} {\bibfnamefont {P.}~\bibnamefont
  {Nordlander}},\ }\href {\doibase 10.1038/nnano.2014.311} {\enquote {\bibinfo
  {title} {{Plasmon-induced hot carrier science and technology}},}\ } (\bibinfo
  {year} {2015})\BibitemShut {NoStop}%
\bibitem [{\citenamefont {Sundararamen}\ \emph {et~al.}(2014)\citenamefont
  {Sundararamen}, \citenamefont {Narang}, \citenamefont {Jermyn}, \citenamefont
  {Goddard~III},\ and\ \citenamefont {Atwater}}]{Sundararaman2014}%
  \BibitemOpen
  \bibfield  {author} {\bibinfo {author} {\bibfnamefont {R.}~\bibnamefont
  {Sundararamen}}, \bibinfo {author} {\bibfnamefont {P.}~\bibnamefont
  {Narang}}, \bibinfo {author} {\bibfnamefont {A.~S.}\ \bibnamefont {Jermyn}},
  \bibinfo {author} {\bibfnamefont {W.~A.}\ \bibnamefont {Goddard~III}}, \ and\
  \bibinfo {author} {\bibfnamefont {H.~A.}\ \bibnamefont {Atwater}},\
  }\href@noop {} {\bibfield  {journal} {\bibinfo  {journal} {Nature
  Communications}\ }\textbf {\bibinfo {volume} {5}},\ \bibinfo {pages} {5788}
  (\bibinfo {year} {2014})}\BibitemShut {NoStop}%
\bibitem [{\citenamefont {Wu}\ \emph {et~al.}(2015)\citenamefont {Wu},
  \citenamefont {Chen}, \citenamefont {McBride},\ and\ \citenamefont
  {Lian}}]{Wu2015}%
  \BibitemOpen
  \bibfield  {author} {\bibinfo {author} {\bibfnamefont {K.}~\bibnamefont
  {Wu}}, \bibinfo {author} {\bibfnamefont {J.}~\bibnamefont {Chen}}, \bibinfo
  {author} {\bibfnamefont {J.~R.}\ \bibnamefont {McBride}}, \ and\ \bibinfo
  {author} {\bibfnamefont {T.}~\bibnamefont {Lian}},\ }\href {\doibase
  10.1126/science.aac5443} {\bibfield  {journal} {\bibinfo  {journal}
  {Science}\ }\textbf {\bibinfo {volume} {349}},\ \bibinfo {pages} {632}
  (\bibinfo {year} {2015})}\BibitemShut {NoStop}%
\bibitem [{\citenamefont {Yu}\ \emph {et~al.}(2018)\citenamefont {Yu},
  \citenamefont {Sundaresan},\ and\ \citenamefont {Willets}}]{Yu2018}%
  \BibitemOpen
  \bibfield  {author} {\bibinfo {author} {\bibfnamefont {Y.}~\bibnamefont
  {Yu}}, \bibinfo {author} {\bibfnamefont {V.}~\bibnamefont {Sundaresan}}, \
  and\ \bibinfo {author} {\bibfnamefont {K.~A.}\ \bibnamefont {Willets}},\
  }\href {\doibase 10.1021/acs.jpcc.7b12080} {\bibfield  {journal} {\bibinfo
  {journal} {Journal of Physical Chemistry C}\ }\textbf {\bibinfo {volume}
  {122}},\ \bibinfo {pages} {5040} (\bibinfo {year} {2018})}\BibitemShut
  {NoStop}%
\bibitem [{\citenamefont {Kozawa}\ \emph {et~al.}(2016)\citenamefont {Kozawa},
  \citenamefont {Carvalho}, \citenamefont {Verzhbitskiy}, \citenamefont
  {Giustiniano}, \citenamefont {Miyauchi}, \citenamefont {Mouri}, \citenamefont
  {{Castro Neto}}, \citenamefont {Matsuda},\ and\ \citenamefont
  {Eda}}]{Kozawa2016}%
  \BibitemOpen
  \bibfield  {author} {\bibinfo {author} {\bibfnamefont {D.}~\bibnamefont
  {Kozawa}}, \bibinfo {author} {\bibfnamefont {A.}~\bibnamefont {Carvalho}},
  \bibinfo {author} {\bibfnamefont {I.}~\bibnamefont {Verzhbitskiy}}, \bibinfo
  {author} {\bibfnamefont {F.}~\bibnamefont {Giustiniano}}, \bibinfo {author}
  {\bibfnamefont {Y.}~\bibnamefont {Miyauchi}}, \bibinfo {author}
  {\bibfnamefont {S.}~\bibnamefont {Mouri}}, \bibinfo {author} {\bibfnamefont
  {A.~H.}\ \bibnamefont {{Castro Neto}}}, \bibinfo {author} {\bibfnamefont
  {K.}~\bibnamefont {Matsuda}}, \ and\ \bibinfo {author} {\bibfnamefont
  {G.}~\bibnamefont {Eda}},\ }\href {\doibase 10.1021/acs.nanolett.6b00801}
  {\bibfield  {journal} {\bibinfo  {journal} {Nano Letters}\ }\textbf {\bibinfo
  {volume} {16}},\ \bibinfo {pages} {4087} (\bibinfo {year}
  {2016})}\BibitemShut {NoStop}%
\bibitem [{\citenamefont {Feigenbaum}\ \emph {et~al.}(2010)\citenamefont
  {Feigenbaum}, \citenamefont {Diest},\ and\ \citenamefont
  {Atwater}}]{Feigenbaum2010}%
  \BibitemOpen
  \bibfield  {author} {\bibinfo {author} {\bibfnamefont {E.}~\bibnamefont
  {Feigenbaum}}, \bibinfo {author} {\bibfnamefont {K.}~\bibnamefont {Diest}}, \
  and\ \bibinfo {author} {\bibfnamefont {H.~A.}\ \bibnamefont {Atwater}},\
  }\href {\doibase 10.1021/nl1006307} {\bibfield  {journal} {\bibinfo
  {journal} {Nano Letters}\ }\textbf {\bibinfo {volume} {10}},\ \bibinfo
  {pages} {2111} (\bibinfo {year} {2010})}\BibitemShut {NoStop}%
\bibitem [{\citenamefont {Folland}\ \emph {et~al.}(2019)\citenamefont
  {Folland}, \citenamefont {Nordin}, \citenamefont {Wasserman},\ and\
  \citenamefont {Caldwell}}]{Folland2019}%
  \BibitemOpen
  \bibfield  {author} {\bibinfo {author} {\bibfnamefont {T.~G.}\ \bibnamefont
  {Folland}}, \bibinfo {author} {\bibfnamefont {L.}~\bibnamefont {Nordin}},
  \bibinfo {author} {\bibfnamefont {D.}~\bibnamefont {Wasserman}}, \ and\
  \bibinfo {author} {\bibfnamefont {J.~D.}\ \bibnamefont {Caldwell}},\ }\href
  {\doibase 10.1063/1.5090777} {\bibfield  {journal} {\bibinfo  {journal}
  {Journal of Applied Physics}\ }\textbf {\bibinfo {volume} {125}},\ \bibinfo
  {pages} {191102} (\bibinfo {year} {2019})}\BibitemShut {NoStop}%
\bibitem [{\citenamefont {Tandon}\ \emph {et~al.}(2019)\citenamefont {Tandon},
  \citenamefont {Agrawal}, \citenamefont {Heo},\ and\ \citenamefont
  {Milliron}}]{Tandon2019}%
  \BibitemOpen
  \bibfield  {author} {\bibinfo {author} {\bibfnamefont {B.}~\bibnamefont
  {Tandon}}, \bibinfo {author} {\bibfnamefont {A.}~\bibnamefont {Agrawal}},
  \bibinfo {author} {\bibfnamefont {S.}~\bibnamefont {Heo}}, \ and\ \bibinfo
  {author} {\bibfnamefont {D.~J.}\ \bibnamefont {Milliron}},\ }\href {\doibase
  10.1021/acs.nanolett.9b00079} {\bibfield  {journal} {\bibinfo  {journal}
  {Nano Letters}\ }\textbf {\bibinfo {volume} {19}},\ \bibinfo {pages} {2012}
  (\bibinfo {year} {2019})}\BibitemShut {NoStop}%
\bibitem [{\citenamefont {Lin}\ \emph {et~al.}(2008)\citenamefont {Lin},
  \citenamefont {Zhigilei},\ and\ \citenamefont {Celli}}]{Lin2008}%
  \BibitemOpen
  \bibfield  {author} {\bibinfo {author} {\bibfnamefont {Z.}~\bibnamefont
  {Lin}}, \bibinfo {author} {\bibfnamefont {L.~V.}\ \bibnamefont {Zhigilei}}, \
  and\ \bibinfo {author} {\bibfnamefont {V.}~\bibnamefont {Celli}},\ }\href
  {\doibase 10.1103/PhysRevB.77.075133} {\bibfield  {journal} {\bibinfo
  {journal} {Physical Review B - Condensed Matter and Materials Physics}\
  }\textbf {\bibinfo {volume} {77}},\ \bibinfo {pages} {1} (\bibinfo {year}
  {2008})}\BibitemShut {NoStop}%
\bibitem [{\citenamefont {Sachet}\ \emph {et~al.}(2015)\citenamefont {Sachet},
  \citenamefont {Shelton}, \citenamefont {Harris}, \citenamefont {Gaddy},
  \citenamefont {Irving}, \citenamefont {Curtarolo}, \citenamefont {Donovan},
  \citenamefont {Hopkins}, \citenamefont {Sharma}, \citenamefont {Sharma},
  \citenamefont {Ihlefeld}, \citenamefont {Franzen},\ and\ \citenamefont
  {Maria}}]{Sachet2015}%
  \BibitemOpen
  \bibfield  {author} {\bibinfo {author} {\bibfnamefont {E.}~\bibnamefont
  {Sachet}}, \bibinfo {author} {\bibfnamefont {C.~T.}\ \bibnamefont {Shelton}},
  \bibinfo {author} {\bibfnamefont {J.~S.}\ \bibnamefont {Harris}}, \bibinfo
  {author} {\bibfnamefont {B.~E.}\ \bibnamefont {Gaddy}}, \bibinfo {author}
  {\bibfnamefont {D.~L.}\ \bibnamefont {Irving}}, \bibinfo {author}
  {\bibfnamefont {S.}~\bibnamefont {Curtarolo}}, \bibinfo {author}
  {\bibfnamefont {B.~F.}\ \bibnamefont {Donovan}}, \bibinfo {author}
  {\bibfnamefont {P.~E.}\ \bibnamefont {Hopkins}}, \bibinfo {author}
  {\bibfnamefont {P.~A.}\ \bibnamefont {Sharma}}, \bibinfo {author}
  {\bibfnamefont {A.~L.}\ \bibnamefont {Sharma}}, \bibinfo {author}
  {\bibfnamefont {J.}~\bibnamefont {Ihlefeld}}, \bibinfo {author}
  {\bibfnamefont {S.}~\bibnamefont {Franzen}}, \ and\ \bibinfo {author}
  {\bibfnamefont {J.-p.}\ \bibnamefont {Maria}},\ }\href {\doibase
  10.1038/NMAT4203} {\bibfield  {journal} {\bibinfo  {journal} {Nature
  Materials}\ }\textbf {\bibinfo {volume} {14}},\ \bibinfo {pages} {414}
  (\bibinfo {year} {2015})}\BibitemShut {NoStop}%
\bibitem [{\citenamefont {Runnerstrom}\ \emph {et~al.}(2017)\citenamefont
  {Runnerstrom}, \citenamefont {Kelley}, \citenamefont {Sachet}, \citenamefont
  {Shelton},\ and\ \citenamefont {Maria}}]{Runnerstrom2017}%
  \BibitemOpen
  \bibfield  {author} {\bibinfo {author} {\bibfnamefont {E.~L.}\ \bibnamefont
  {Runnerstrom}}, \bibinfo {author} {\bibfnamefont {K.~P.}\ \bibnamefont
  {Kelley}}, \bibinfo {author} {\bibfnamefont {E.}~\bibnamefont {Sachet}},
  \bibinfo {author} {\bibfnamefont {C.~T.}\ \bibnamefont {Shelton}}, \ and\
  \bibinfo {author} {\bibfnamefont {J.~P.}\ \bibnamefont {Maria}},\ }\href
  {\doibase 10.1021/acsphotonics.7b00429} {\bibfield  {journal} {\bibinfo
  {journal} {ACS Photonics}\ }\textbf {\bibinfo {volume} {4}},\ \bibinfo
  {pages} {1885} (\bibinfo {year} {2017})}\BibitemShut {NoStop}%
\bibitem [{\citenamefont {Runnerstrom}\ \emph {et~al.}(2019)\citenamefont
  {Runnerstrom}, \citenamefont {Kelley}, \citenamefont {Folland}, \citenamefont
  {Nolen}, \citenamefont {Engheta}, \citenamefont {Caldwell},\ and\
  \citenamefont {Maria}}]{Runnerstrom2019}%
  \BibitemOpen
  \bibfield  {author} {\bibinfo {author} {\bibfnamefont {E.~L.}\ \bibnamefont
  {Runnerstrom}}, \bibinfo {author} {\bibfnamefont {K.~P.}\ \bibnamefont
  {Kelley}}, \bibinfo {author} {\bibfnamefont {T.~G.}\ \bibnamefont {Folland}},
  \bibinfo {author} {\bibfnamefont {J.~R.}\ \bibnamefont {Nolen}}, \bibinfo
  {author} {\bibfnamefont {N.}~\bibnamefont {Engheta}}, \bibinfo {author}
  {\bibfnamefont {J.~D.}\ \bibnamefont {Caldwell}}, \ and\ \bibinfo {author}
  {\bibfnamefont {J.~P.}\ \bibnamefont {Maria}},\ }\href {\doibase
  10.1021/acs.nanolett.8b04182} {\bibfield  {journal} {\bibinfo  {journal}
  {Nano Letters}\ }\textbf {\bibinfo {volume} {19}},\ \bibinfo {pages} {948}
  (\bibinfo {year} {2019})}\BibitemShut {NoStop}%
\bibitem [{\citenamefont {Choi}\ \emph {et~al.}(2014)\citenamefont {Choi},
  \citenamefont {Wilson},\ and\ \citenamefont {Cahill}}]{Choi2014}%
  \BibitemOpen
  \bibfield  {author} {\bibinfo {author} {\bibfnamefont {G.-M.}\ \bibnamefont
  {Choi}}, \bibinfo {author} {\bibfnamefont {R.~B.}\ \bibnamefont {Wilson}}, \
  and\ \bibinfo {author} {\bibfnamefont {D.~G.}\ \bibnamefont {Cahill}},\
  }\href@noop {} {\bibfield  {journal} {\bibinfo  {journal} {Physical Review
  B}\ }\textbf {\bibinfo {volume} {89}},\ \bibinfo {pages} {64307} (\bibinfo
  {year} {2014})}\BibitemShut {NoStop}%
\bibitem [{\citenamefont {Radue}\ \emph {et~al.}(2018)\citenamefont {Radue},
  \citenamefont {Tomko}, \citenamefont {Giri},\ and\ \citenamefont
  {Braun}}]{Radue2018}%
  \BibitemOpen
  \bibfield  {author} {\bibinfo {author} {\bibfnamefont {E.}~\bibnamefont
  {Radue}}, \bibinfo {author} {\bibfnamefont {J.}~\bibnamefont {Tomko}},
  \bibinfo {author} {\bibfnamefont {A.}~\bibnamefont {Giri}}, \ and\ \bibinfo
  {author} {\bibfnamefont {J.~L.}\ \bibnamefont {Braun}},\ }\href {\doibase
  10.1021/acsphotonics.8b01045} {\bibfield  {journal} {\bibinfo  {journal} {ACS
  Photonics}\ }\textbf {\bibinfo {volume} {5}},\ \bibinfo {pages} {4880}
  (\bibinfo {year} {2018})}\BibitemShut {NoStop}%
\bibitem [{\citenamefont {Qiu}\ \emph {et~al.}(1994)\citenamefont {Qiu},
  \citenamefont {Juhasz}, \citenamefont {Suarez}, \citenamefont {Bron},\ and\
  \citenamefont {Tien}}]{Qiu1994a}%
  \BibitemOpen
  \bibfield  {author} {\bibinfo {author} {\bibfnamefont {T.}~\bibnamefont
  {Qiu}}, \bibinfo {author} {\bibfnamefont {T.}~\bibnamefont {Juhasz}},
  \bibinfo {author} {\bibfnamefont {C.}~\bibnamefont {Suarez}}, \bibinfo
  {author} {\bibfnamefont {W.}~\bibnamefont {Bron}}, \ and\ \bibinfo {author}
  {\bibfnamefont {C.}~\bibnamefont {Tien}},\ }\href {\doibase
  10.1016/0017-9310(94)90397-2} {\bibfield  {journal} {\bibinfo  {journal}
  {International Journal of Heat and Mass Transfer}\ }\textbf {\bibinfo
  {volume} {37}},\ \bibinfo {pages} {2799} (\bibinfo {year}
  {1994})}\BibitemShut {NoStop}%
\bibitem [{\citenamefont {Qiu}\ and\ \citenamefont {Tien}(1994)}]{Qiu1994}%
  \BibitemOpen
  \bibfield  {author} {\bibinfo {author} {\bibfnamefont {T.}~\bibnamefont
  {Qiu}}\ and\ \bibinfo {author} {\bibfnamefont {C.}~\bibnamefont {Tien}},\
  }\href {\doibase 10.1016/0017-9310(94)90396-4} {\bibfield  {journal}
  {\bibinfo  {journal} {International Journal of Heat and Mass Transfer}\
  }\textbf {\bibinfo {volume} {37}},\ \bibinfo {pages} {2789} (\bibinfo {year}
  {1994})}\BibitemShut {NoStop}%
\bibitem [{\citenamefont {Lyeo}\ and\ \citenamefont {Cahill}(2006)}]{Lyeo2006}%
  \BibitemOpen
  \bibfield  {author} {\bibinfo {author} {\bibfnamefont {H.-k.}\ \bibnamefont
  {Lyeo}}\ and\ \bibinfo {author} {\bibfnamefont {D.~G.}\ \bibnamefont
  {Cahill}},\ }\href {\doibase 10.1103/PhysRevB.73.144301} {\bibfield
  {journal} {\bibinfo  {journal} {Physical Review B}\ }\textbf {\bibinfo
  {volume} {73}},\ \bibinfo {pages} {1} (\bibinfo {year} {2006})}\BibitemShut
  {NoStop}%
\bibitem [{\citenamefont {Giri}\ and\ \citenamefont
  {Hopkins}(2019)}]{Giri2019}%
  \BibitemOpen
  \bibfield  {author} {\bibinfo {author} {\bibfnamefont {A.}~\bibnamefont
  {Giri}}\ and\ \bibinfo {author} {\bibfnamefont {P.~E.}\ \bibnamefont
  {Hopkins}},\ }\href {\doibase 10.1002/adfm.201903857} {\bibfield  {journal}
  {\bibinfo  {journal} {Advanced Functional Materials}\ }\textbf {\bibinfo
  {volume} {2019}},\ \bibinfo {pages} {1903857} (\bibinfo {year}
  {2019})}\BibitemShut {NoStop}%
\bibitem [{\citenamefont {Yu}\ \emph {et~al.}(2019)\citenamefont {Yu},
  \citenamefont {Wijesekara}, \citenamefont {Xi},\ and\ \citenamefont
  {Willets}}]{Yu2019}%
  \BibitemOpen
  \bibfield  {author} {\bibinfo {author} {\bibfnamefont {Y.}~\bibnamefont
  {Yu}}, \bibinfo {author} {\bibfnamefont {K.~D.}\ \bibnamefont {Wijesekara}},
  \bibinfo {author} {\bibfnamefont {X.}~\bibnamefont {Xi}}, \ and\ \bibinfo
  {author} {\bibfnamefont {K.~A.}\ \bibnamefont {Willets}},\ }\href {\doibase
  10.1021/acsnano.9b00219} {\bibfield  {journal} {\bibinfo  {journal} {ACS
  Nano}\ }\textbf {\bibinfo {volume} {13}},\ \bibinfo {pages} {3629} (\bibinfo
  {year} {2019})}\BibitemShut {NoStop}%
\bibitem [{\citenamefont {Ratchford}\ \emph {et~al.}(2017)\citenamefont
  {Ratchford}, \citenamefont {Dunkelberger}, \citenamefont {Vurgaftman},
  \citenamefont {Owrutsky},\ and\ \citenamefont {Pehrsson}}]{Ratchford2017}%
  \BibitemOpen
  \bibfield  {author} {\bibinfo {author} {\bibfnamefont {D.~C.}\ \bibnamefont
  {Ratchford}}, \bibinfo {author} {\bibfnamefont {A.~D.}\ \bibnamefont
  {Dunkelberger}}, \bibinfo {author} {\bibfnamefont {I.}~\bibnamefont
  {Vurgaftman}}, \bibinfo {author} {\bibfnamefont {C.}~\bibnamefont
  {Owrutsky}}, \ and\ \bibinfo {author} {\bibfnamefont {P.~E.}\ \bibnamefont
  {Pehrsson}},\ }\href {\doibase 10.1021/acs.nanolett.7b02366} {\bibfield
  {journal} {\bibinfo  {journal} {Nano Letters}\ }\textbf {\bibinfo {volume}
  {17}},\ \bibinfo {pages} {6047} (\bibinfo {year} {2017})}\BibitemShut
  {NoStop}%
\bibitem [{\citenamefont {Giri}\ \emph {et~al.}(2015)\citenamefont {Giri},
  \citenamefont {Gaskins}, \citenamefont {Donovan}, \citenamefont
  {Szwejkowski}, \citenamefont {Warzoha}, \citenamefont {Rodriguez},
  \citenamefont {Ihlefeld},\ and\ \citenamefont {Hopkins}}]{Giri2015}%
  \BibitemOpen
  \bibfield  {author} {\bibinfo {author} {\bibfnamefont {A.}~\bibnamefont
  {Giri}}, \bibinfo {author} {\bibfnamefont {J.~T.}\ \bibnamefont {Gaskins}},
  \bibinfo {author} {\bibfnamefont {B.~F.}\ \bibnamefont {Donovan}}, \bibinfo
  {author} {\bibfnamefont {C.}~\bibnamefont {Szwejkowski}}, \bibinfo {author}
  {\bibfnamefont {R.~J.}\ \bibnamefont {Warzoha}}, \bibinfo {author}
  {\bibfnamefont {M.~A.}\ \bibnamefont {Rodriguez}}, \bibinfo {author}
  {\bibfnamefont {J.}~\bibnamefont {Ihlefeld}}, \ and\ \bibinfo {author}
  {\bibfnamefont {P.~E.}\ \bibnamefont {Hopkins}},\ }\href {\doibase
  10.1063/1.4914867} {\bibfield  {journal} {\bibinfo  {journal} {Journal of
  Applied Physics}\ }\textbf {\bibinfo {volume} {117}} (\bibinfo {year}
  {2015}),\ 10.1063/1.4914867}\BibitemShut {NoStop}%
\bibitem [{\citenamefont {Zhou}\ \emph {et~al.}(2019)\citenamefont {Zhou},
  \citenamefont {Tokina}, \citenamefont {Tomko}, \citenamefont {Braun},
  \citenamefont {Hopkins},\ and\ \citenamefont {Prezhdo}}]{Zhou2019}%
  \BibitemOpen
  \bibfield  {author} {\bibinfo {author} {\bibfnamefont {X.}~\bibnamefont
  {Zhou}}, \bibinfo {author} {\bibfnamefont {M.~V.}\ \bibnamefont {Tokina}},
  \bibinfo {author} {\bibfnamefont {J.~A.}\ \bibnamefont {Tomko}}, \bibinfo
  {author} {\bibfnamefont {J.~L.}\ \bibnamefont {Braun}}, \bibinfo {author}
  {\bibfnamefont {P.~E.}\ \bibnamefont {Hopkins}}, \ and\ \bibinfo {author}
  {\bibfnamefont {O.~V.}\ \bibnamefont {Prezhdo}},\ }\href {\doibase
  10.1063/1.5096901} {\bibfield  {journal} {\bibinfo  {journal} {Journal of
  Chemical Physics}\ }\textbf {\bibinfo {volume} {150}},\ \bibinfo {pages}
  {184701} (\bibinfo {year} {2019})}\BibitemShut {NoStop}%
\bibitem [{\citenamefont {Yang}\ \emph {et~al.}(2017)\citenamefont {Yang},
  \citenamefont {Kelley}, \citenamefont {Sachet}, \citenamefont {Campione},
  \citenamefont {Luk}, \citenamefont {Maria}, \citenamefont {Sinclair},\ and\
  \citenamefont {Brener}}]{Yang2017}%
  \BibitemOpen
  \bibfield  {author} {\bibinfo {author} {\bibfnamefont {Y.}~\bibnamefont
  {Yang}}, \bibinfo {author} {\bibfnamefont {K.}~\bibnamefont {Kelley}},
  \bibinfo {author} {\bibfnamefont {E.}~\bibnamefont {Sachet}}, \bibinfo
  {author} {\bibfnamefont {S.}~\bibnamefont {Campione}}, \bibinfo {author}
  {\bibfnamefont {T.~S.}\ \bibnamefont {Luk}}, \bibinfo {author} {\bibfnamefont
  {J.-P.}\ \bibnamefont {Maria}}, \bibinfo {author} {\bibfnamefont {M.~B.}\
  \bibnamefont {Sinclair}}, \ and\ \bibinfo {author} {\bibfnamefont
  {I.}~\bibnamefont {Brener}},\ }\href {\doibase 10.1038/nphoton.2017.64}
  {\bibfield  {journal} {\bibinfo  {journal} {Nature Photonics}\ }\textbf
  {\bibinfo {volume} {11}},\ \bibinfo {pages} {390} (\bibinfo {year}
  {2017})}\BibitemShut {NoStop}%
\bibitem [{\citenamefont {Liu}\ \emph {et~al.}(2016)\citenamefont {Liu},
  \citenamefont {Foo}, \citenamefont {Kamruzzaman}, \citenamefont {Ho},
  \citenamefont {Zapien}, \citenamefont {Zhu}, \citenamefont {Li},
  \citenamefont {Walukiewicz},\ and\ \citenamefont {Yu}}]{Liu2016}%
  \BibitemOpen
  \bibfield  {author} {\bibinfo {author} {\bibfnamefont {C.~P.}\ \bibnamefont
  {Liu}}, \bibinfo {author} {\bibfnamefont {Y.}~\bibnamefont {Foo}}, \bibinfo
  {author} {\bibfnamefont {M.}~\bibnamefont {Kamruzzaman}}, \bibinfo {author}
  {\bibfnamefont {C.~Y.}\ \bibnamefont {Ho}}, \bibinfo {author} {\bibfnamefont
  {J.~A.}\ \bibnamefont {Zapien}}, \bibinfo {author} {\bibfnamefont
  {W.}~\bibnamefont {Zhu}}, \bibinfo {author} {\bibfnamefont {Y.~J.}\
  \bibnamefont {Li}}, \bibinfo {author} {\bibfnamefont {W.}~\bibnamefont
  {Walukiewicz}}, \ and\ \bibinfo {author} {\bibfnamefont {K.~M.}\ \bibnamefont
  {Yu}},\ }\href {\doibase 10.1103/PhysRevApplied.6.064018} {\bibfield
  {journal} {\bibinfo  {journal} {Physical Review Applied}\ }\textbf {\bibinfo
  {volume} {6}},\ \bibinfo {pages} {064018} (\bibinfo {year}
  {2016})}\BibitemShut {NoStop}%
\bibitem [{\citenamefont {Nolen}\ \emph {et~al.}(2019)\citenamefont {Nolen},
  \citenamefont {Runnerstrom}, \citenamefont {Kelley}, \citenamefont {Luk},
  \citenamefont {Folland}, \citenamefont {Cleri}, \citenamefont {Maria},\ and\
  \citenamefont {Caldwell}}]{Nolen2019}%
  \BibitemOpen
  \bibfield  {author} {\bibinfo {author} {\bibfnamefont {J.~R.}\ \bibnamefont
  {Nolen}}, \bibinfo {author} {\bibfnamefont {E.~L.}\ \bibnamefont
  {Runnerstrom}}, \bibinfo {author} {\bibfnamefont {K.~P.}\ \bibnamefont
  {Kelley}}, \bibinfo {author} {\bibfnamefont {T.~S.}\ \bibnamefont {Luk}},
  \bibinfo {author} {\bibfnamefont {T.~G.}\ \bibnamefont {Folland}}, \bibinfo
  {author} {\bibfnamefont {A.}~\bibnamefont {Cleri}}, \bibinfo {author}
  {\bibfnamefont {J.-P.}\ \bibnamefont {Maria}}, \ and\ \bibinfo {author}
  {\bibfnamefont {J.~D.}\ \bibnamefont {Caldwell}},\ }\href@noop {} {\bibfield
  {journal} {\bibinfo  {journal} {Physical Review Materials}\ }\textbf
  {\bibinfo {volume} {4}},\ \bibinfo {pages} {025202} (\bibinfo {year}
  {2019})}\BibitemShut {NoStop}%
\bibitem [{\citenamefont {Giannozzi}\ \emph {et~al.}(2009)\citenamefont
  {Giannozzi}, \citenamefont {Baroni}, \citenamefont {Bonini}, \citenamefont
  {Calandra}, \citenamefont {Car}, \citenamefont {Cavazzoni}, \citenamefont
  {Ceresoli}, \citenamefont {Chiarotti}, \citenamefont {Cococcioni},
  \citenamefont {Dabo}, \citenamefont {Corso}, \citenamefont {Gironcoli},
  \citenamefont {Gerstmann}, \citenamefont {Gougoussis}, \citenamefont
  {Kokalj}, \citenamefont {Lazzeri}, \citenamefont {Martin-samos},
  \citenamefont {Marzari}, \citenamefont {Mauri}, \citenamefont {Mazzarello},
  \citenamefont {Paolini}, \citenamefont {Pasquarello}, \citenamefont
  {Paulatto},\ and\ \citenamefont {Sbraccia}}]{Giannozzi2009}%
  \BibitemOpen
  \bibfield  {author} {\bibinfo {author} {\bibfnamefont {P.}~\bibnamefont
  {Giannozzi}}, \bibinfo {author} {\bibfnamefont {S.}~\bibnamefont {Baroni}},
  \bibinfo {author} {\bibfnamefont {N.}~\bibnamefont {Bonini}}, \bibinfo
  {author} {\bibfnamefont {M.}~\bibnamefont {Calandra}}, \bibinfo {author}
  {\bibfnamefont {R.}~\bibnamefont {Car}}, \bibinfo {author} {\bibfnamefont
  {C.}~\bibnamefont {Cavazzoni}}, \bibinfo {author} {\bibfnamefont
  {D.}~\bibnamefont {Ceresoli}}, \bibinfo {author} {\bibfnamefont {G.~L.}\
  \bibnamefont {Chiarotti}}, \bibinfo {author} {\bibfnamefont {M.}~\bibnamefont
  {Cococcioni}}, \bibinfo {author} {\bibfnamefont {I.}~\bibnamefont {Dabo}},
  \bibinfo {author} {\bibfnamefont {A.~D.}\ \bibnamefont {Corso}}, \bibinfo
  {author} {\bibfnamefont {S.~D.}\ \bibnamefont {Gironcoli}}, \bibinfo {author}
  {\bibfnamefont {U.}~\bibnamefont {Gerstmann}}, \bibinfo {author}
  {\bibfnamefont {C.}~\bibnamefont {Gougoussis}}, \bibinfo {author}
  {\bibfnamefont {A.}~\bibnamefont {Kokalj}}, \bibinfo {author} {\bibfnamefont
  {M.}~\bibnamefont {Lazzeri}}, \bibinfo {author} {\bibfnamefont
  {L.}~\bibnamefont {Martin-samos}}, \bibinfo {author} {\bibfnamefont
  {N.}~\bibnamefont {Marzari}}, \bibinfo {author} {\bibfnamefont
  {F.}~\bibnamefont {Mauri}}, \bibinfo {author} {\bibfnamefont
  {R.}~\bibnamefont {Mazzarello}}, \bibinfo {author} {\bibfnamefont
  {S.}~\bibnamefont {Paolini}}, \bibinfo {author} {\bibfnamefont
  {A.}~\bibnamefont {Pasquarello}}, \bibinfo {author} {\bibfnamefont
  {L.}~\bibnamefont {Paulatto}}, \ and\ \bibinfo {author} {\bibfnamefont
  {C.}~\bibnamefont {Sbraccia}},\ }\href {\doibase
  10.1088/0953-8984/21/39/395502} {\bibfield  {journal} {\bibinfo  {journal}
  {J. Phys.: Condens. Matter}\ }\textbf {\bibinfo {volume} {21}},\ \bibinfo
  {pages} {395502} (\bibinfo {year} {2009})}\BibitemShut {NoStop}%
\bibitem [{\citenamefont {Gianozzi}\ \emph {et~al.}(2017)\citenamefont
  {Gianozzi}, \citenamefont {Andreussi}, \citenamefont {Brumme}, \citenamefont
  {Bunau}, \citenamefont {Buongiorno~Nardelli}, \citenamefont {Calandra},
  \citenamefont {Car}, \citenamefont {Cavazzoni}, \citenamefont {Ceresoli},\
  and\ \citenamefont {Cococcioni}}]{Gianozzi2017}%
  \BibitemOpen
  \bibfield  {author} {\bibinfo {author} {\bibfnamefont {P.}~\bibnamefont
  {Gianozzi}}, \bibinfo {author} {\bibfnamefont {O.}~\bibnamefont {Andreussi}},
  \bibinfo {author} {\bibfnamefont {T.}~\bibnamefont {Brumme}}, \bibinfo
  {author} {\bibfnamefont {O.}~\bibnamefont {Bunau}}, \bibinfo {author}
  {\bibfnamefont {M.}~\bibnamefont {Buongiorno~Nardelli}}, \bibinfo {author}
  {\bibfnamefont {M.}~\bibnamefont {Calandra}}, \bibinfo {author}
  {\bibfnamefont {R.}~\bibnamefont {Car}}, \bibinfo {author} {\bibfnamefont
  {C.}~\bibnamefont {Cavazzoni}}, \bibinfo {author} {\bibfnamefont
  {D.}~\bibnamefont {Ceresoli}}, \ and\ \bibinfo {author} {\bibfnamefont
  {M.}~\bibnamefont {Cococcioni}},\ }\href@noop {} {\bibfield  {journal}
  {\bibinfo  {journal} {Journal of Physics: Condensed Matter}\ }\textbf
  {\bibinfo {volume} {29}},\ \bibinfo {pages} {465901} (\bibinfo {year}
  {2017})}\BibitemShut {NoStop}%
\bibitem [{\citenamefont {Aras}\ and\ \citenamefont
  {Kılı{\c{c}}}(2014)}]{Aras2014}%
  \BibitemOpen
  \bibfield  {author} {\bibinfo {author} {\bibfnamefont {M.}~\bibnamefont
  {Aras}}\ and\ \bibinfo {author} {\bibfnamefont {C.}~\bibnamefont
  {Kılı{\c{c}}}},\ }\href {\doibase 10.1063/1.4890458} {\bibfield  {journal}
  {\bibinfo  {journal} {The Journal of Chemical Physics}\ }\textbf {\bibinfo
  {volume} {141}},\ \bibinfo {pages} {044106} (\bibinfo {year}
  {2014})}\BibitemShut {NoStop}%
\bibitem [{\citenamefont {Goh}\ \emph {et~al.}(2017)\citenamefont {Goh},
  \citenamefont {Mah},\ and\ \citenamefont {Yoon}}]{Goh2017}%
  \BibitemOpen
  \bibfield  {author} {\bibinfo {author} {\bibfnamefont {E.~S.}\ \bibnamefont
  {Goh}}, \bibinfo {author} {\bibfnamefont {J.~W.}\ \bibnamefont {Mah}}, \ and\
  \bibinfo {author} {\bibfnamefont {T.~L.}\ \bibnamefont {Yoon}},\ }\href
  {\doibase 10.1016/j.commatsci.2017.06.032} {\bibfield  {journal} {\bibinfo
  {journal} {Computational Materials Science}\ }\textbf {\bibinfo {volume}
  {138}},\ \bibinfo {pages} {111} (\bibinfo {year} {2017})}\BibitemShut
  {NoStop}%
\bibitem [{\citenamefont {Akimov}\ and\ \citenamefont
  {Prezhdo}(2013)}]{Akimov2013}%
  \BibitemOpen
  \bibfield  {author} {\bibinfo {author} {\bibfnamefont {A.~V.}\ \bibnamefont
  {Akimov}}\ and\ \bibinfo {author} {\bibfnamefont {O.~V.}\ \bibnamefont
  {Prezhdo}},\ }\href {\doibase 10.1021/ct400641n} {\bibfield  {journal}
  {\bibinfo  {journal} {J. Chem. Theory Comput.}\ }\textbf {\bibinfo {volume}
  {9}},\ \bibinfo {pages} {4959} (\bibinfo {year} {2013})}\BibitemShut
  {NoStop}%
\bibitem [{\citenamefont {Akimov}\ and\ \citenamefont
  {Prezhdo}(2014)}]{Akimov2014}%
  \BibitemOpen
  \bibfield  {author} {\bibinfo {author} {\bibfnamefont {A.~V.}\ \bibnamefont
  {Akimov}}\ and\ \bibinfo {author} {\bibfnamefont {O.~V.}\ \bibnamefont
  {Prezhdo}},\ }\href {\doibase 10.1021/ct400934c} {\bibfield  {journal}
  {\bibinfo  {journal} {J. Chem. Theory Comput.}\ }\textbf {\bibinfo {volume}
  {10}},\ \bibinfo {pages} {789} (\bibinfo {year} {2014})}\BibitemShut
  {NoStop}%
\end{thebibliography}%

\section*{Acknowledgments}
We acknowledge funding from the U.S. Department of Defense, Multidisciplinary University Research Initiative through the Army Research Office, Grant No. W911NF-16-1-0406.

\section*{Author Contributions}
P.E.H. and J.-P.M. conceived the idea and supervised the project; J.A.T. performed the TDTR and IR pump-probe measurements and corresponding analysis; E.L.R., K.P.K., J.N., and A.C. grew and characterized the CdO films; Y.-S.W. and O.V.P. performed the \textit{ab-initio} calculations; J.A.T. and D.H.O. performed the TTM calculations; J.A.T., J.R.N. and J.D.C. performed the TMM calculations and provided insight on the dispersion relations. J.A.T. wrote the manuscript with input from all the authors.
\clearpage

 \begin{figure*}
	\centering
	\includegraphics[scale=0.625]{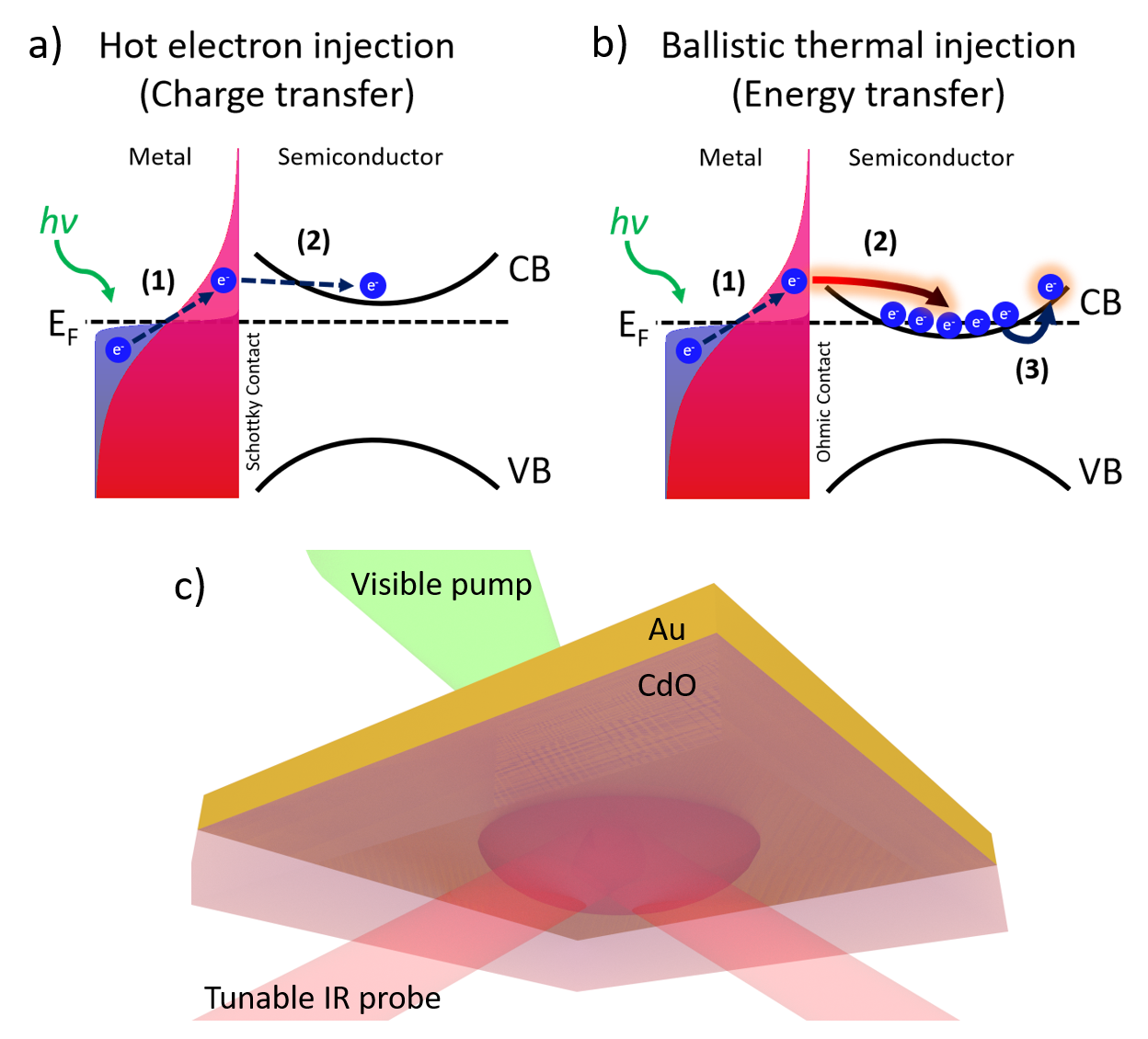}
	\caption{a) The typically-assumed process occurring at metal/semiconductor interfaces following photo-excitation of the metallic contact. In this case, (1) hot-electrons are first generated in the gold. At sufficiently high electron temperatures, (2) the electrons traverse the interface and add charge to the conduction band of the semiconductor. b) Our proposed process for metal/semiconductor interfaces following ultrafast excitation of the metal contact. This mechanism relies on hot-electron generation in the metal (1); prior to electron-phonon coupling (less than a couple picoseconds), energy propagates ballistically towards the metal/semiconductor interface. (2) The electron energy front reaches the interface, whereby the electrons transfer their \textit{energy}, rather than charge, to the pre-existing free electrons in the semiconductor's conduction band. (3) The pre-existing semiconductor's electrons are now at an elevated temperature and are promoted to elevated states in the conduction band (e.g., intraband excitations). c) Schematic of our ultrafast ENZ experiment to spatially resolve the electron energy distribution following potential injection processes. The 520 nm pump beam excites the Au surface at the Au/air interface, while a sub-picosecond probe pulse monitors the ENZ mode of a thin Y:CdO film.}
	\label{Figure1}
\end{figure*}

\begin{figure*}
	\centering
	\includegraphics[scale=0.625]{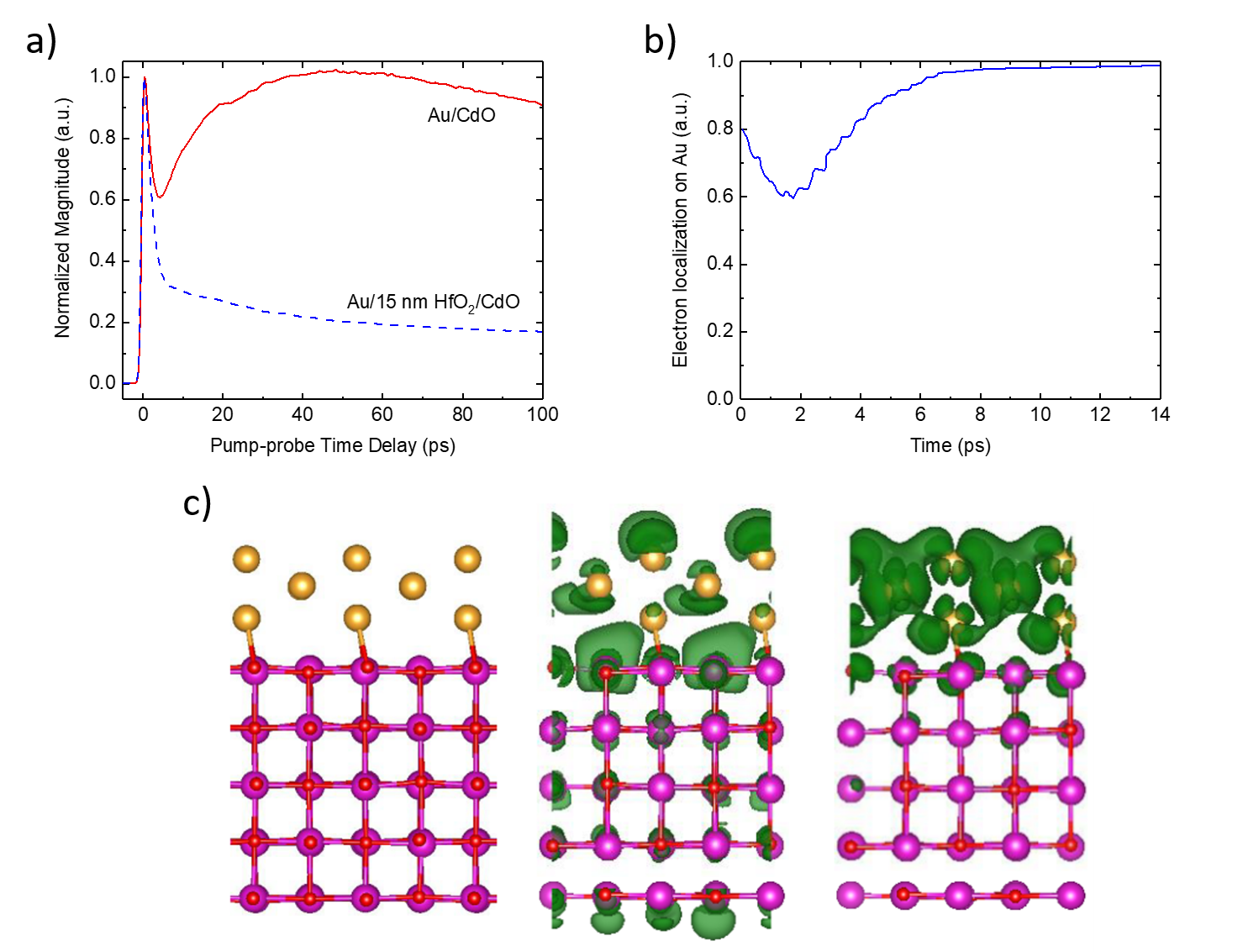}
	\caption{a) TDTR curves for 15 nm Au on 100 nm CdO, with a carrier concentration of $7.7\times10^{19}$ cm$^{-3}$. In the case where the two are in direct contact (straight red line), back-heating is clearly observed, indicative of electron injection. Conversely, the addition of a thin dielectric layer between the two media (dashed blue line) inhibits this injection effect and leads to electron thermalization only within the Au film.  b) Evolution of electron localization on Au atoms; zero-time corresponds to the initially-excited state. While the electron remains primarily localized within the Au slab, its wavefunction extends into the CdO, which allows for a greatly-increased energy transfer rate due to the high frequency modes available within the CdO. After a few picoseconds, as the structure relaxes, the electron re-localizes solely within the Au layer, thus ending the BTI process. c) Left: Optimized structure of the Au/CdO simulation cell. Yellow, large purple and small red balls represent Au, Cd and O atoms respectively. The CdO slab is about twice thicker than the Au slab, mimicking experiments. Middle and Right: Charge densities of the initial pumped and final states, respectively. The excited hot electron localized on Au has a tail into CdO.  The relaxed electron is localized nearly fully on Au, because the Au Fermi level is inside CdO bandgap.}
	\label{Figure2}
\end{figure*}

\begin{figure*}
	\centering
	\includegraphics[scale=0.5]{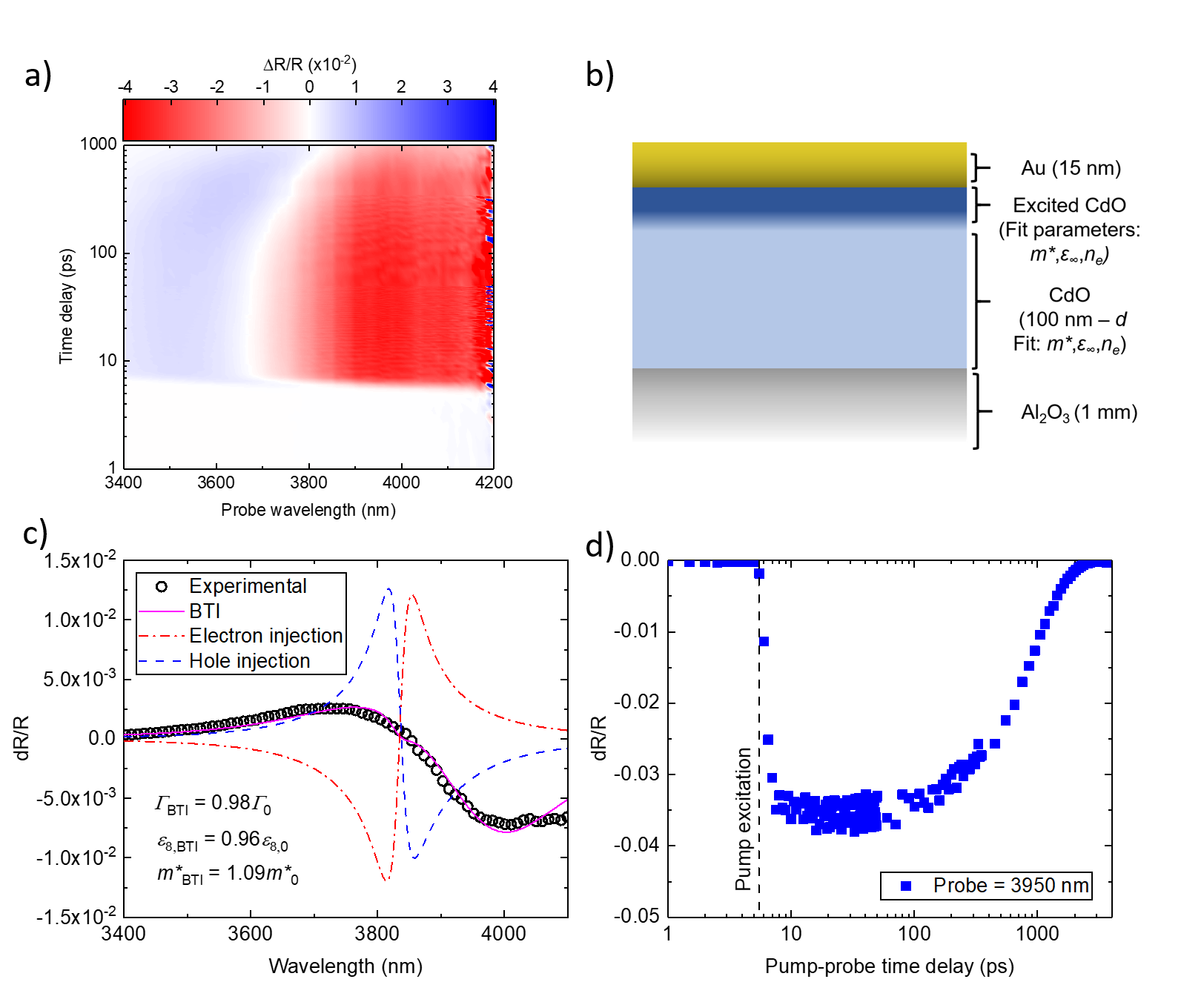}
	\caption{a) Transient reflectivity measurements at a fluence of $\sim$0.5 J m$^{-2}$. The x-axis denote the wavelength of the probe beam, while the y-axis reprsents the time delay between the 520 nm pump pulse and the tunable IR probe beam; the spectral resolution of these data is 10 nanometers. Note, the pump pulse arrives at $\sim$6 picoseconds. Additionally, red represents a \textit{decrease} in probe reflectance, while the blue represents an increased reflectance. b) Schematic of the TMM simulation along with the various fit parameters. c) Change in reflectivity for the Au/CdO heterostructure immediately following pump exictation and the simulated reflectance curves calculated via transfer matrix method for various injection mechanisms. Clearly, neither hole or electron injection are able to capture the observed trend. The inset values are the TMM best-fit values, which determined a 5 nm perturbation layer in agreement with our two-temperature model calculations.  d) Transient reflectivity of the Au/CdO heterostructure probed at 3950 nm following visible excitation of the Au film; the 1/$e$ decay time of this curve is approximately 700 picoseconds.}
	\label{Figure3}
\end{figure*}

\end{document}